\documentclass[aps,prb,superscriptaddress,floatfix,showpacs,twocolumn]{revtex4}
\usepackage{amsfonts,amsmath,amssymb,bm}
\usepackage{graphicx,graphics,float}
\usepackage{color}
\usepackage[colorlinks,citecolor=blue,linkcolor=blue]{hyperref}

\begin{document}
%\begin{CJK*}{GBK}{}
\title{Thermoelectric transport with electron-phonon coupling and
electron-electron interaction in molecular junctions}

\author{Jie Ren}\thanks{Present address: Theoretical Division, 
Los Alamos National Laboratory,
Los Alamos, New Mexico 87545, USA.}
%\email{renjie@nus.edu.sg}
\affiliation{NUS Graduate School for Integrative Sciences and
Engineering, Singapore 117456, Republic of Singapore}
\affiliation{Department of Physics and Centre for Computational
Science and Engineering, National University of Singapore, Singapore
117546, Republic of Singapore}

\author{Jian-Xin Zhu}%\email{jxzhu@lanl.gov}
\affiliation{Theoretical Division, Los Alamos National Laboratory,
Los Alamos, New Mexico 87545, USA}

\author{James E. Gubernatis}
\affiliation{Theoretical Division, Los Alamos National Laboratory,
Los Alamos, New Mexico 87545, USA}

\author{Chen Wang}
\affiliation{Department of Physics and Centre
for Computational Science and Engineering, National University of
Singapore, Singapore 117546, Republic of Singapore}
\affiliation{Department of Physics, Zhejiang University, Hangzhou
310027, P. R. China}

\author{Baowen Li}
\affiliation{NUS Graduate School for Integrative Sciences and
Engineering, Singapore 117456, Republic of Singapore}
\affiliation{Department of Physics and Centre for Computational
Science and Engineering, National University of Singapore, Singapore
117546, Republic of Singapore}

\date{\today}
\begin{abstract}
Within the framework of nonequilibrium Green's functions,  we
investigate the thermoelectric transport in a single molecular
junction with electron-phonon and electron-electron interactions. By transforming into a displaced phonon basis, we are able to deal with these interactions nonperturbatively. Then, by invoking the weak tunneling limit, we are able to calculate the thermoelectricity. Results show that at low
temperatures, resonances of the thermoelectric figure of merit, $ZT$,
occur around the sides of resonances of electronic conductance but
drop dramatically to zero at exactly these resonant points. We find $ZT$ can be enhanced by increasing electron-phonon coupling and Coulomb
repulsion, and an optimal enhancement is obtained when
these two interactions are competing. Our results indicate a great potential for
single molecular junctions as good thermoelectric devices over a
wide range of temperatures.
\end{abstract}

%73.63.Kv Quantum dots ((low rank)73.63.-b Electronic transport in nanoscale materials and structures)
%65.80.-g Thermal properties of small particles, nanocrystals, nanotubes, and other related systems
%72.15.Jf Thermoelectric and thermomagnetic effects
%73.23.Hk Coulomb blockade; single-electron tunneling
%85.65.+h Molecular electronic devices
%71.38.-k Polarons and electron-phonon interactions

%72.10.Di Scattering by phonons, magnons, and other nonlocalized excitations
%73.23.-b Electronic transport in mesoscopic systems
%73.63.Rt Nanoscale contacts

 % 03.65.Yz Decoherence; open systems; quantum statistical methods (see also 03.67.Pp in quantum information; for decoherence in Bose-Einstein condensates, see 03.75.Gg)
 % 03.65.Vf Phases: geometric; dynamic or topological
 % 05.60.-k Transport processes
 % 05.60.Gg Quantum transport
 % 05.70.Ln  nonequilibrium thermodynamics
 % 44.10.+i Heat conduction (see also 66.60.+a and 66.70.-f in transport properties of condensed matter)
 % 63.20.Ry Anharmonic lattice modes,
 % 63.22.-m Phonons or vibrational states in low-dimensional structures and nanoscale materials
 % 66.70.-f Nonelectronic thermal conduction and heat-pulse propagation in solids; thermal waves (for thermal conduction in metals and alloys, see 72.15.Cz and 72.15.Eb)
\maketitle
%\end{CJK*}

%The interest in developing efficient thermoelectric materials for
%the generation and harvesting of energy spans decades.
\section{Introduction}
Recently, the potential afforded by nanoscale
engineering~\cite{Snyder08} has revitalized interest in
developing novel thermoelectric materials for the generation and
harvesting of energy. It is well accepted that nanoscale materials
engineering in principle creates unlimited opportunities for the
creation of more efficient energy-conversion
devices~\cite{Dresselhaus07} and thus expands the
potential of using thermoelectricity for meeting the challenge of being a sustainable energy
source.~\cite{Majumdar04} However, questions remain about the best
ways for manipulating the microscopic properties of the material so that
enhanced performance occurs.~\cite{Majumdar04,Mahan96}
%Here we address these questions for a simple single molecular junction.

The thermoelectric performance is typically characterized by the
figure of merit, $ZT$,~\cite{Mahan96,Dubi11} which is defined as $ZT
= {G_{e}S^{2}T}/{\kappa}$, where $G_{e}$ is the electronic
conductance, $S$ is the thermopower, %{\color{red} $T$ is the temperature,}
$T$ is the temperature, and $\kappa$ is the thermal
conductance. Increasing the value of $ZT$ increases the efficiency
of heat-electricity conversion.
%~\cite{Snyder08}~\cite{Mahan96,}.
The dependence of the figure of merit on both charge and energy
transport shows that thermoelectric efficiency is strongly affected
by the underlying electronic and vibrational properties of a
material. These dependencies are especially transparent in molecular
junctions,~\cite{Reddy07} as charge accumulation on the junction
causes Coulomb interactions (e-e) to perturb the electronic
structure~\cite{ee02} and the electron-phonon (e-ph) coupling to
perturb the vibrational modes and the conformation of the
junction.~\cite{Galperin07}

The energy scale of e-e interaction is usually much larger than that
of e-ph interaction. However, at the atomic and molecular levels, the
electrodes can screen the Coulomb repulsion, reducing it to the same
order of magnitude as the e-ph interaction. The interactions now
compete. Thus, it is of fundamental and practical importance to
explore the effects of this competition to gain insights into the
optimization of thermoelectric transport for better energy-conversion
devices.

In this work, we investigate the thermoelectric transport in a
single molecular junction with e-e interaction and e-ph coupling of
arbitrary strengths in the framework of the nonequilibrium
Green's function (NEGF) method.~\cite{Haugbook} Although there are considerable efforts toward understanding the effects of the e-e~\cite{Murphy08,
Costi10, Liu10, Wysokinski10} or e-ph~\cite{Wohlman10,Zianni10} interactions,
what happens to the thermoelectric transport when e-ph coupling competes
with e-e interaction in a full range of strength is still an open
question. %This is the major question we address. In particular,
In contrast to previous
work in the literature,~\cite{Koch04,Leijnse10} we treat the e-e and e-ph interactions within the molecular junction nonperturbatively, by a transform of the phonon basis with effective displacements. This treatment moves our use of the NEGF framework beyond the weak e-ph coupling perturbative
analysis,~\cite{eph09, Wohlman10, Paulsson05} the strong e-ph
coupling limit of canonical transformation
(the Lang-Firsov approach),~\cite{Zhu03Chen05, Mahanbook, Kuo10} and the mean-field
approximation in the strong Coulomb repulsion regime.~\cite{Liu10,
Kuo10}

%As a consequence of strong e-e and e-ph competition, the thermoelectric coefficient can become significantly enhanced.

%An important consequence of treating the molecular junction
%interactions exactly is our currents in and out of the junction are
%{\em conserved}, a property often lacking in approximate treatments.

%\textbf{What new effects are seen as a consequences of this exact approach and treating e-e and e-ph together?}\varnothing

\section{Method and Approximation}

We start with the standard Anderson-Holstein Hamiltonian:~\cite{AHM1,AHM2}
%\begin{equation*}
$H=H_\text{mol}+H_\text{T}+H_\text{leads}$.
%\end{equation*}
$H_\text{mol}$ describes the molecular junction of one orbital
level, with Coulomb repulsion between electrons of opposite spin orientations and additional coupling to the vibration of itself, which is conventionally
assumed to be:~\cite{Haugbook,Koch04,Leijnse10,Kuo10,Mahanbook,AHM1,AHM2}
%\begin{equation}
\begin{multline*}
H_\text{mol}=\omega_0\hat{a}^{\dag}\hat{a}+\sum_{\sigma}\varepsilon_{\sigma}\hat{d}^{\dag}_{\sigma}\hat{d}^{}_{\sigma} \\
+\sum_{\sigma}\lambda_{\sigma}\hat{d}^{\dag}_{\sigma}\hat{d}^{}_{\sigma}(\hat{a}^{\dag}
+\hat{a})+U\hat{d}^{\dag}_{\uparrow}\hat{d}_{\uparrow}\hat{d}^{\dag}_{\downarrow}\hat{d}_{\downarrow}.
\end{multline*}
%\end{equation}
In the first term on the right-hand side, $\hat{a}^{\dag}$ and
$\hat{a}$ create and annihilate a phonon with energy $\omega_0$
while in the second term $\hat{d}^{\dag}_{\sigma}$ and
$\hat{d}_{\sigma}$ create and annihilate an electron of spin $\sigma$
at the molecular level with energies
$\varepsilon_{\uparrow}=\varepsilon_0+\Delta\varepsilon/2$ or
$\varepsilon_{\downarrow}=\varepsilon_0-\Delta\varepsilon/2$.
%$\varepsilon_{\uparrow(\downarrow)}=\varepsilon_0\pm\Delta\varepsilon/2$.
The third and fourth terms describe the e-ph and e-e interactions with
strengths $\lambda_{\sigma}$ and $U$. The expression
\begin{equation*}
H_{\text{leads}}=\sum_{k,\sigma,\nu=\text{L,R}}\epsilon^{\nu}_{k\sigma}N^{\nu}_{k\sigma}.
\end{equation*}
describes the left and right
electrode leads with $N^{\nu}_{k\sigma}=\hat{c}^{\dag}_{k\sigma\nu}\hat{c}^{}_{k\sigma\nu}$ the number operator for electrons of reservoir $\nu$ with wave number
$k$ and spin $\sigma$. The expressoin
\begin{equation*}
H_{\text{T}}=
\sum_{k,\sigma,\nu=\text{L,R}}t^{\nu}_{k\sigma}
(\hat{d}^{\dag}_{\sigma}\hat{c}^{}_{k\sigma\nu}+
 \hat{d}^{}_{\sigma}\hat{c}^{\dag}_{k\sigma\nu})
\end{equation*}
describes the tunneling Hamiltonian of the electron hopping between the molecule and the electrode leads. Here, due to the
large mismatch of vibrational spectra between the molecule and
metallic electrodes, the phonon transport is not considered.

Computing the transport for this model within the NEGF formalism requires the knowledge of various Green's functions for the different parts of $H$. We start by analytically solving the eigenproblem for the molecular part of the junction. %finding these quantities for the junction.
The Hilbert space of the molecular part is spanned by the basis
\{$|\vartheta, n\rangle$\}, where $|\vartheta\rangle$ are the four possible electron states
$|\o\rangle$, $|\sigma\rangle$, $|\bar\sigma\rangle$,
$|\sigma\bar{\sigma}\rangle$ and $|n\rangle$ denotes phonon states with $n=0,1,\ldots,\infty$. To nonperturbatively treat the e-ph and e-e interactions
with arbitrary strengths, we first block diagonalize the Hamiltonian
with respect to electron states: %with elements:
\begin{eqnarray*}
\langle\o|H_{\mathrm{mol}}|\o\rangle&=&\omega_0\hat{a}^{\dag}\hat{a}, \\
\langle\sigma|H_{\mathrm{mol}}|\sigma\rangle&=&\omega_0\hat{a}^{\dag}\hat{a}
+\varepsilon_{\sigma}+\lambda_{\sigma}(\hat{a}^{\dag}+\hat{a}), \\
\langle\sigma\bar{\sigma}|H_{\mathrm{mol}}|\sigma\bar{\sigma}\rangle&=& \omega_0\hat{a}^{\dag}\hat{a}
+\sum_{\sigma}\varepsilon_{\sigma}+\sum_{\sigma}\lambda_{\sigma}(\hat{a}^{\dag}+\hat{a})+U. \end{eqnarray*}
%$\langle\o|H_{\mathrm{mol}}|\o\rangle=\omega_0\hat{a}^{\dag}\hat{a}$,
%$\langle\sigma|H_{\mathrm{mol}}|\sigma\rangle=\omega_0\hat{a}^{\dag}\hat{a}
%+\varepsilon_{\sigma}+\lambda_{\sigma}(\hat{a}^{\dag}+\hat{a})$,
%$\langle\sigma\bar{\sigma}|H_{\mathrm{mol}}|\sigma\bar{\sigma}\rangle= \omega_0\hat{a}^{\dag}\hat{a}
%+\sum_{\sigma}\varepsilon_{\sigma}+\sum_{\sigma}\lambda_{\sigma}(\hat{a}^{\dag}+\hat{a})+U$
It is easy to further diagonalize
$\langle\o|H_{\mathrm{mol}}|\o\rangle$ with conventional Fock phonon
states:
$|n\rangle_{\o}=|n\rangle=[(\hat{a}^{\dag})^n/\sqrt{n!}]|0\rangle$.
However, in order to diagonalize the other two matrix elements, we need introduce a new phonon basis,~\cite{Chen08, Chen11} with displacements shifted by different electron states through the e-ph coupling:
\begin{equation}
|n\rangle_{\vartheta}=[{(\hat{A}^{\dag}_{\vartheta})^n}/{\sqrt{n!}}]
\exp{(-g_{\vartheta}^2/2-g_{\vartheta}\hat{a}^{\dag})}|0\rangle,
\end{equation}
where $\hat{A}^{\dag}_{\vartheta}=\hat{a}^{\dag}+g_{\vartheta}$ denotes the new creator that creates a phonon displaced from the original position by
a value $g_{\vartheta}$ depending on the electronic state, that is,
$g_{\o}=0$, $g_{\sigma}=\lambda_{\sigma}/\omega_0$, and
$g_{\sigma\bar\sigma}=g_{\sigma}+g_{\bar\sigma}$. Clearly, when electrons are absent on the molecular quantum dot, $g_{\o}=0$ and the displaced phonon basis reduces to the normal Fock state of the phonons. We then can write
\begin{eqnarray*}
%\langle\o|H_{\mathrm{mol}}|\o\rangle&=&\omega_0\hat{A}_{\o}^{\dag}\hat{A}_{\o},\\
\langle\sigma|H_{\mathrm{mol}}|\sigma\rangle&=&\omega_0\hat{A}_{\sigma}^{\dag}\hat{A}_{\sigma}
+\varepsilon_{\sigma}-\omega_0g^2_{\sigma},\\
\langle\sigma\bar{\sigma}|H_{\mathrm{mol}}|\sigma\bar{\sigma}\rangle&=&\omega_0\hat{A}_{\sigma\bar\sigma}^{\dag}\hat{A}_{\sigma\bar\sigma}
+\varepsilon_{\sigma}+
\varepsilon_{\bar{\sigma}}-\omega_0g^2_{\sigma\bar{\sigma}}+U.
\end{eqnarray*}
%$\langle\o|H_{\mathrm{mol}}|\o\rangle=\omega_0\hat{A}_{\o}^{\dag}\hat{A}_{\o}$,
%$\langle\sigma|H_{\mathrm{mol}}|\sigma\rangle=\omega_0\hat{A}_{\sigma}^{\dag}\hat{A}_{\sigma}
%+\varepsilon_{\sigma}-\omega_0g^2_{\sigma}$,
%$\langle\sigma\bar{\sigma}|H_{\mathrm{mol}}|\sigma\bar{\sigma}\rangle=\omega_0\hat{A}_{\sigma\bar\sigma}^{\dag}\hat{A}_{\sigma\bar\sigma}
%+\varepsilon_{\sigma}+
%\varepsilon_{\bar{\sigma}}-\omega_0g^2_{\sigma\bar{\sigma}}+U$.
Therefore, with the help of the new phonon basis, the solution to the eigenvalue
problem is
\begin{eqnarray}
H_{\mathrm{mol}}|\o,n\rangle_{\o}&=&n\omega_0|\o,n\rangle_{\o},\\
H_{\mathrm{mol}}|\sigma,n\rangle_{\sigma}&=&(n\omega_0+\tilde\varepsilon_{\sigma})|\sigma,n\rangle_{\sigma},\\
H_{\mathrm{mol}}|\sigma\bar{\sigma},n\rangle_{\sigma\bar{\sigma}}&=&(n\omega_0+\tilde\varepsilon_{\sigma\bar{\sigma}}+U)|\sigma\bar{\sigma},n\rangle_{\sigma\bar{\sigma}},
\end{eqnarray}
where
%$\tilde\varepsilon_{\sigma} =\varepsilon_{\sigma}-\omega_0g^2_{\sigma}$ and
%$\tilde\varepsilon_{\sigma\bar{\sigma}}=
%\varepsilon_{\sigma}+\varepsilon_{\bar\sigma}-\omega_0g^2_{\sigma\bar{\sigma}}
%=\tilde\varepsilon_{\sigma}+\tilde\varepsilon_{\bar\sigma}
%-2\omega_0g_{\sigma}g_{\bar\sigma}$.
\begin{align*}
\tilde\varepsilon_{\sigma} &=\varepsilon_{\sigma}-\omega_0g^2_{\sigma} \\
\tilde\varepsilon_{\sigma\bar{\sigma}} &=
\varepsilon_{\sigma}+\varepsilon_{\bar\sigma}-\omega_0g^2_{\sigma\bar{\sigma}}
=\tilde\varepsilon_{\sigma}+\tilde\varepsilon_{\bar\sigma}
-2\omega_0g_{\sigma}g_{\bar\sigma}.
\end{align*}
The negative term
$-2\omega_0g_{\sigma}g_{\bar\sigma}$ evidences the attractive
interaction between different electron states induced by the e-ph
coupling.

With $H_\text{mol}$ diagonalized, we can now analytically calculate the
advanced and retarded Green's functions of the molecule, which are found to be (see Appendix~\ref{appendixA} for details)
%$G^>_{\sigma}(t)=-i\langle d_{\sigma}(t)d^{\dag}_{\sigma}(0)\rangle$ in the frequency domain.
\begin{multline}
G_{\sigma}^{r(a)}(\omega)=\frac{1}{Z}\sum^{\infty}_{n,m=0}\left[\frac{e^{-\beta
m\omega_0}+e^{-\beta
(n\omega_0+\tilde\varepsilon_{\sigma})}}{\omega-\Delta_{mn}^{(1)}\pm
i0^{+}} + \right.\\
\left.\frac{e^{-\beta
(m\omega_0+\tilde\varepsilon_{\bar\sigma})}+e^{-\beta
(n\omega_0+\tilde\varepsilon_{\sigma\bar\sigma}+U)}}{\omega-\Delta_{mn}^{(2)}\pm
i0^{+}}\right]D^2_{nm}(g_{\sigma}),
\label{eq:G}
\end{multline}
where
\begin{align*}
\Delta_{nm}^{(1)} &=(n-m)\omega_0+\tilde\varepsilon_{\sigma}, \\
\Delta_{nm}^{(2)} &=(n-m)\omega_0+(\tilde\varepsilon_{\sigma}-2\omega_0 g_{\sigma}g_{\bar{\sigma}}+U), \\
D_{nm}(g_{\sigma})&=(-1)^m {{}_{\o}} \langle m|n \rangle_\sigma
 =(-1)^m {}_{\bar\sigma}\langle  m|n\rangle_{\sigma{\bar\sigma}} \\
 &=e^{-g_{\sigma}^2/2}\sum^{\mathrm{min}\{n,m\}}_{k=0}\frac{(-1)^k\sqrt{n!m!}g_{\sigma}^{n+m-2k}}{(n-k)!(m-k)!k!}, \\
%\begin{align*}\label{poles}
%\Delta_{nm}^{(1)}&=(n-m)\omega_0+\tilde\varepsilon_{\sigma}, \\
%\Delta_{nm}^{(2)}&=(n-m)\omega_0+(\tilde\varepsilon_{\sigma}-2\omega_0 g_{\sigma}g_{\bar{\sigma}}+U), \\
%D_{nm}(g_{\sigma})&=(-1)^m {{}_{\o}} \langle m|n \rangle_\sigma
%=(-1)^m {}_{\bar\sigma}\langle  m|n\rangle_{\sigma{\bar\sigma}} \\
%&=e^{-g_{\sigma}^2/2}\sum^{\mathrm{min}\{n,m\}}_{k=0}\frac{(-1)^k\sqrt{n!m!}g_{\sigma}^{n+m-2k}}{(n-k)!(m-k)!k!},
%\end{align*}
Z &=(1+N_{\mathrm{ph}})(1+e^{-\beta \tilde\varepsilon_{\sigma}}+e^{-\beta
\tilde\varepsilon_{\bar\sigma}}+e^{-\beta(\tilde\varepsilon_{\sigma\bar\sigma}+U)}).
\end{align*}
Here $N_{\mathrm{ph}}=1/(e^{\beta \omega_0}-1)$
denotes the Bose distribution of the phonon population with inverse temperature
$\beta\equiv k_BT$. Note, in the zero-temperature limit, Eq.~(\ref{eq:G}) is consistent with the ``atomic limit'' Green's functions in Refs.~\onlinecite{AHM2,Martin08} through using canonical transformation. The advantage of our method is that at finite temperatures our results still give Green's functions analytically and explicitly, while for the atomic limit with the Lang-Firsov canonical transformation, the expressions of Green's functions contain the electronic level occupation, the value of which must be found through a self-consistency iteration.

With the molecular part treated nonperturbatively, we now follow standard
paths to build the Green's functions of the molecule-lead-electrode system. By using the Dyson equation and the Keldysh
formula,~\cite{Haugbook,Mahanbook} we have the total retarded
(advanced) Green's function $$G^{r(a)}_{\mathrm{tot},\sigma}=[(G^{r(a)}_{\sigma})^{-1}-\Sigma^{r(a)}_{\mathrm{lead},\sigma}]^{-1}$$
and the total lesser (greater) Green's function $$G^{<(>)}_{\mathrm{tot},\sigma}=
G^r_{\mathrm{tot},\sigma}\Sigma^{<(>)}_{\mathrm{tot},\sigma}G^a_{\mathrm{tot},\sigma}.$$
The total self-energy has two contributions:
$\Sigma_{\mathrm{tot},\sigma}=\Sigma_{\mathrm{lead},\sigma}+\Sigma_{\mathrm{int},\sigma}$, where $\Sigma_{\mathrm{int},\sigma}$ is the contribution from the e-ph and e-e interactions, following from \cite{Haugbook,Mahanbook}
%\begin{align}
$\Sigma^{r(a)}_{\mathrm{int},\sigma} ={G^{r(a)}_{0,\sigma}}^{-1}-{G^{r(a)}_{\sigma}}^{-1}$, %\nonumber\\
and $\Sigma^{<(>)}_{\mathrm{int},\sigma}
={G^{r}_{\sigma}}^{-1}G^{<(>)}_{\sigma}{G^{a}_{\sigma}}^{-1}$,
%\end{align}
where $G^{r(a)}_{0,\sigma}(\omega)=(\omega-\varepsilon_\sigma\pm
i0^{+})^{-1}$ denotes the noninteracting Green's functions without involving e-ph and e-e interactions. Here $\Sigma_{\mathrm{lead},\sigma}$ depicts the contribution from two tunneling parts between the molecular quantum dot and leads.

So far, no approximations have been made for the representation. However, to exactly obtain $\Sigma_{\mathrm{lead},\sigma}$ in a strong correlated system is highly nontrivial.  In the following, we take the lowest order approximation of $\Sigma_{\mathrm{lead},\sigma}$ as in a noninteracting system and obtain~\cite{Haugbook}
\begin{align*}
\Sigma^{>}_{\mathrm{lead},\sigma} &=-i[\Gamma_{\sigma}^L(1-f_L)+\Gamma_{\sigma}^R(1-f_R)], \\
\Sigma^{<}_{\mathrm{lead},\sigma} &=i(\Gamma_{\sigma}^Lf_L+\Gamma_{\sigma}^Rf_R), \\
\Sigma^{r(a)}_{\mathrm{lead},\sigma} &=\mp i\Gamma_{\sigma},
\end{align*}
where $\Gamma_{\sigma}\equiv{(\Gamma_{\sigma}^L+\Gamma_{\sigma}^R)}/{2}$,
$\Gamma^{\nu}_{\sigma}(\omega) \equiv 2\pi\sum_k|t^{\nu}_{k\sigma}|^2
\delta(\omega-\epsilon^{\nu}_{k\sigma})$, denoting the molecule-electrode coupling functions, are energy independent in
the wideband limit, and $f_{\nu}=[e^{\beta_{\nu}(\omega-\mu_{\nu})}+1]^{-1}$ are the Fermi-Dirac distributions of two reservoirs. This approximation is valid in the weak tunneling limit; i.e., we take small values of $\Gamma^{\nu}_{\sigma}$, comparing to all other energy scales. This weak tunneling approximation is consistent with the polaron tunneling approximation in Ref.~\onlinecite{Maier11}, under the Lang-Firsov canonical transformation.
Note this weak tunneling approximation also ignores the Kondo effect, which is justified since we are interested in regimes beyond the Kondo temperature. An interpolative approach for self-energies~\cite{Martin08} may be used to relax this weak tunneling limit.

With the Green's functions in hand,
we can now study the quantum transport by calculating the charge current
\begin{equation*}
J_{\nu}=-e\langle \sum_{k\sigma}\dot N^{\nu}_{k\sigma}\rangle
 \end{equation*}
and heat current
\begin{equation*}
I^{\nu}_Q=-\langle\sum_{k\sigma}(\epsilon^{\nu}_{k\sigma}-\mu_{\nu})\dot
N^{\nu}_{k\sigma}\rangle,
 \end{equation*}
leaving electrode $\nu$, which in terms of the total Green's functions are~\cite{Haugbook}
\begin{align*}
J_{\nu} &=\frac{i e}{h}\sum_{\sigma}\int
d\omega\left(\Gamma^{\nu}_{\sigma}G^<_{\mathrm{tot},\sigma}+f_{\nu}\Gamma^{\nu}_{\sigma}[G^r_{\mathrm{tot},\sigma}-G^a_{\mathrm{tot},\sigma}]\right), \\
I^{\nu}_Q &=\frac{i}{h}\sum_{\sigma}\int d\omega(\omega-\mu_\nu)\left(\Gamma^{\nu}_{\sigma}G^<_{\mathrm{tot},\sigma}+f_{\nu}\Gamma^{\nu}_{\sigma} [G^r_{\mathrm{tot},\sigma}-G^a_{\mathrm{tot},\sigma}]\right).
\end{align*}
%$J_{\nu}=-e\langle \sum_{k\sigma}\dot N^{\nu}_{k\sigma}\rangle$
%and heat current
%$I^{\nu}_Q=-\langle\sum_{k\sigma}(\epsilon^{\nu}_{k\sigma}-\mu_{\nu})\dot
%N^{\nu}_{k\sigma}\rangle$
%leaving electrode $\nu$, which in terms of the total Green's functions are \cite{Haugbook}
%$J_{\nu} =\frac{i e}{h}\sum_{\sigma}\int
%d\omega\left(\Gamma^{\nu}_{\sigma}G^<_{\mathrm{tot},\sigma}+f_{\nu}\Gamma^{\nu}_{\sigma}[G^r_{\mathrm{tot},\sigma}-G^a_{\mathrm{tot},\sigma}]\right)$ and
%$I^{\nu}_Q =\frac{i}{h}\sum_{\sigma}\int d\omega(\omega-\mu_\nu)\left(\Gamma^{\nu}_{\sigma}G^<_{\mathrm{tot},\sigma}+f_{\nu}\Gamma^{\nu}_{\sigma} [G^r_{\mathrm{tot},\sigma}-G^a_{\mathrm{tot},\sigma}]\right)$.
After some algebra (see details in Appendix~\ref{appendixB}), we find that the currents going from the left electrode to the central system are
\begin{align}
J_L &=\frac{e}{\hbar}\int\frac{d\omega}{2\pi}\mathcal{T}(\omega)\big[f_L(\omega)-f_R(\omega)\big],\label{eq:current} \\
I^L_Q
&=\frac{1}{\hbar}\int\frac{d\omega}{2\pi}(\omega-\mu_L)\mathcal{T}(\omega)\big[f_L(\omega)-f_R(\omega)\big],\label{eq:heatcurrent}
\end{align}
where
\begin{equation*}
\mathcal{T}(\omega)=\mathcal{T}_{\uparrow}(\omega)+\mathcal{T}_{\downarrow}(\omega)
\end{equation*}
with
\begin{equation}
\mathcal{T}_{\sigma}(\omega)=\frac{\Gamma^L_{\sigma}\Gamma^R_{\sigma}}{[(G^{r}_{\sigma})^{-1}+i\Gamma_{\sigma}][(G^{a}_{\sigma})^{-1}-i\Gamma_{\sigma}]}.
\end{equation}
We can also obtain the same expressions for $J_R$ and $I^R_Q$ with $L\leftrightarrow R$, depicting the electron and heat current from the junction to the right electrode.

Any meaningful transport theory in terms of NEGF must respect current conservation:~\cite{Haugbook}
\begin{equation*}
J_L+J_R=\frac{e}{\hbar}\int\frac{d\omega}{2\pi}\mathrm{Tr}\left[\Sigma^>_{\mathrm{int}}G^<_{\mathrm{tot}}
-\Sigma^<_{\mathrm{int}}G^>_{\mathrm{tot}}\right]=0.
\end{equation*}
The self-consistent Born approximation is an example of a conserving approximation. Without the self-consistency of Green's functions and self-energies, the current nonconserving issue $J_L\neq -J_R$ generally exists, for example, in the perturbation approximation of weak e-ph coupling~\cite{Lu07} and the canonical transformation for strong e-ph coupling.~\cite{Zhu03Chen05}
People usually choose to calculate the symmetrized
current $J\doteq(J_L-J_R)/2$ to avoid the current nonconserving issue. The present approach, however, satisfies $J_L=-J_R$ directly. %thus solves the
%issue in approximate treatments of e-ph coupling.
It also avoids the time-consuming self-consistent refinement in the
mean-field treatment of Coulomb repulsion.~\cite{Haugbook}

Energy conservation also gives the relation
$-\omega_0\dot N_{\mathrm{ph}}=I^L_Q+I^R_Q$.~\cite{Wohlman10} For linear response, the system is in equilibrium such that $\dot N_{\mathrm{ph}}=0$ and $I^L_Q=-I^R_Q$, since $N_{\mathrm{ph}}$ follows the Bose
distribution for given temperature $T$. Therefore, the equilibrium temperature is directly used for calculation, and indeed our heat currents satisfy $I^L_Q=-I^R_Q$. %For the nonlinear response case, the system is in non-equilibrium and $I^{L(R)}_Q$ contains the effective inverse
%temperature $\beta_{\mathrm{eff}}$ which is a function of $N_{\mathrm{ph}}$:
%$\beta_{\mathrm{eff}}=\frac{1}{\omega_0}\ln\frac{1+N_{\mathrm{ph}}}{N_{\mathrm{ph}}}$.
%Thus, to obtain the phonon population and effective temperature, we can iteratively obtain the
%steady state solution of $0=-\omega_0\dot N_{\mathrm{ph}}=I^L_Q+I^R_Q$ which
%gives the non-equilbirium $N_{\mathrm{ph}}$ and $\beta_{\mathrm{eff}}$, and in
%turn gives non-equilibrium currents and other transport properties.

\section{Thermoelectric transport and Discussions}
The thermoelectric coefficients are conventionally considered around
the linear response region:~\cite{Mahanbook}
\begin{align*}
\mu_{L(R)}&=\mu_F\pm e\Delta V/2, \\
T_{L(R)}&=T\pm\Delta T/2,
\end{align*}
which yields
\begin{eqnarray*}
\left(%
\begin{array}{c}
  J \\
  I_Q \\
\end{array}%
\right)=\left(%
\begin{array}{cc}
  G_e & G_eST \\
  G_eST & (G_eS^2T+\kappa_e)T \\
\end{array}%
\right)
\left(%
\begin{array}{c}
  \Delta V \\
  {\Delta T}/{T} \\
\end{array}
\right),
\end{eqnarray*}
where
\begin{gather}
G_{e} = \left.J/\Delta V\right|_{\Delta T=0} =e^2\mathcal{L}_0, \label{Ge}\\
S= \left.-{\Delta V}/{\Delta T}\right|_{J=0} ={\mathcal{L}_1}/{(eT\mathcal{L}_0)}, \label{S}\\
\kappa_{e} =\left.{I_{Q}}/{\Delta T}\right|_{J=0}=(\mathcal{L}_2-{\mathcal{L}^2_1}/{\mathcal{L}_0})/T, \label{ke}
\end{gather}
with
%$\mathcal{L}_n=({1}/{h})\int
%d\omega\mathcal{T}(\omega)(\omega-\mu_F)^n\left(-{\partial
%f}/{\partial\omega}\right)$.
\begin{equation}
\mathcal{L}_n=\frac{1}{h}\int
d\omega\mathcal{T}(\omega)(\omega-\mu_F)^n\left(-\frac{\partial
f}{\partial\omega}\right)\;.
\label{Ln}
\end{equation}
While $G_eS^2T+\kappa_e$ is the thermal
conductance at zero voltage bias, $\kappa_e$ is the conventional
thermal conductance of electrons at zero electron current. Another
important quantity is the Lorenz ratio $L=\kappa_e/(G_eT)$. For
macroscopic conductors, the Wiedemann-Franz (WF) law relates the
electronic and heat conductances via the universal relation
$L=L_0\equiv(\pi^2/3)(k_B/e)^2$, which indicates that charge and
energy currents suffer from the same scattering mechanisms such that
more electrons carry more heat and vice versa. In a
single-electron transistor, the Coulomb blockade effect leads to the
strong violation of WF law:\ $L/L_0\gg1$.~\cite{Kubala08} However,
in order to obtain a large $ZT={G_eS^2T}/{\kappa_e} =S^2/L$,
the opposite violation of WF law, $L/L_0\ll1$, is desirable.

%%%%%%%%%%%%%%%%%%%%%%%%%%%%%%%%%%%%%%%%%%%%%%%%%%%%%%%%%%%%%%%%%%%%%%%%%%%%%%%%%%%%%%%%%%%
\begin{figure}%[!t]
%\begin{center}
\scalebox{0.42}[0.4]{\includegraphics{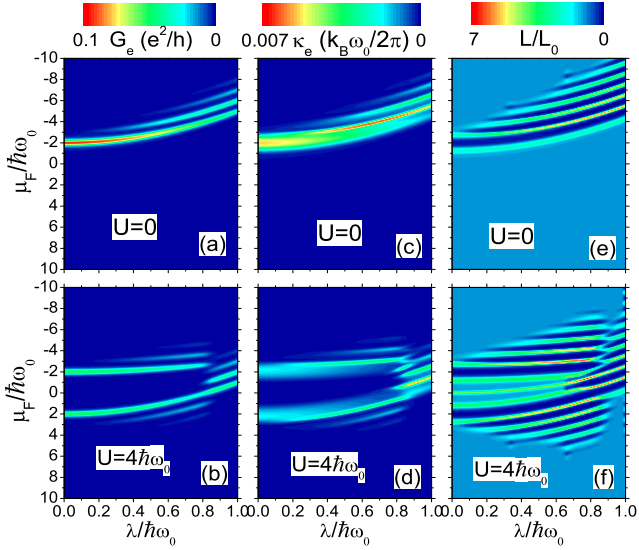}}\vspace{-3mm}
%\centering\includegraphics[width=1.0\linewidth]{L.eps}
\caption{(Color online) Electronic and thermal conductances and
Lorenz ratio as a function of $\mu_F$ for different e-ph and e-e
interaction strengths: $\varepsilon_0=-2\hbar\omega_0$,
$\Delta\varepsilon=0$, $\Gamma^L_{\sigma}=\Gamma^R_{\sigma}=0.05\hbar\omega_0$,
$\hbar\omega_0=30$ meV, and $T=35$ K, which are typical experimental
values.~\cite{Paulsson05}} \label{L}
%\end{center}
\end{figure}
%%%%%%%%%%%%%%%%%%%%%%%%%%%%%%%%%%%%%%%%%%%%%%%%%%%%%%%%%%%%%%%%%%%%%%%%%%%%%%%%%%%%%%%%%%%

\begin{figure}%[!htb]
%\vspace{4.5cm}
\scalebox{0.34}[0.30]{\includegraphics{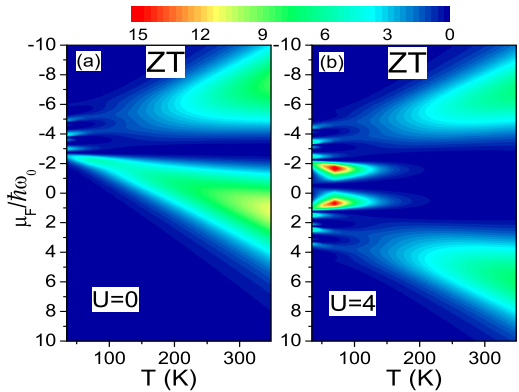}}\vspace{-3mm}
%\centering\includegraphics[width=1.0\linewidth]{ZT.pdf}
\caption{(Color online) Temperature-dependence of $ZT$ for moderate
$\lambda=0.5\hbar\omega_0$, with (a) $U=0$ and (b)
$U=4\hbar\omega_0$. Other parameters are the same as those in
Fig.~\ref{L}.}\label{T}
\end{figure}

Figures~\ref{L}(a) and~\ref{L}(b) show the electronic conductance $G_e$ as a
function of $\mu_F$ for different values of $\lambda$ and $U$. We
see that the number of resonance peaks increases as the e-ph
coupling increases: Large $\lambda$ excites more phonons and enables
multi-phonon-assisted tunneling. At low temperature, we can approximate $G_e$ as $\mathcal T(\mu_F)$ since $-\partial f/\partial \omega\approx\delta(\omega-\mu_F)$. Therefore, from Eqs.~(\ref{eq:G}) and~(\ref{T}), the positions of the resonance
peaks are determined by the poles of Green's functions; i.e., $\mu_F$ of resonance positions are equal to
$\Delta_{nm}^{(1)}$ or $\Delta_{nm}^{(2)}$. However, exponential
functions and $D_{nm}$ in Eq.~(\ref{eq:G}) weight the resonances and
make the peaks unobservable for certain parameter ranges. From
Eq.~(\ref{eq:G}), when $U$ is so weak that
$\tilde\varepsilon_{\sigma\bar\sigma}+U<\tilde\varepsilon_{\sigma}<0$,
$\Delta^{(2)}_{nm}<0$ thus dominates the resonance peaks
[see Fig.~\ref{L}(a)]. While $U$ increases up to
$\tilde\varepsilon_{\sigma\bar\sigma}+U>\tilde\varepsilon_{\sigma}$,
two resonance branches, $\Delta^{(2)}_{nm}>0$ and
$\Delta^{(1)}_{nm}<0$, emerge [see Fig.~\ref{L}(b)]. For large $U$, when $\lambda$
increases further such that again
$\tilde\varepsilon_{\sigma\bar\sigma}+U<\tilde\varepsilon_{\sigma}$,
then $\Delta^{(2)}_{nm}<0$ redominates the resonances and two
resonant branches merge together, as shown in Fig.~\ref{L}(b). It is a consequence of the competition between the e-ph coupling and e-e interaction with the molecular quantum dot junction.

In Figs.~\ref{L}(a) and~\ref{L}(b), we also observe that
increasing e-ph coupling decreases the peak value of the main resonance of $G_e$
but increases the values of side peaks of $G_e$. This occurs because $D^2_{00}=e^{-g^2}$ decreases with $g$ increasing, while for a positive integer $n$,
$D^2_{0n}=g^{2n}e^{-g^2}/n!$ has the opposite tendency at $g\in[0,\sqrt n]$. When e-ph coupling $g$ increases further ($g>\sqrt n$), $D^2_{0n}$ decreases again so that side peak values of $G_e$ will be repressed by the strong e-ph scattering.  Comparing Figs.~\ref{L}(a) and~\ref{L}(b), we further see that for the weak and moderate
$\lambda$, increasing the Coulomb repulsion reduces $G_e$, which is a
consequence of the factor $e^{-\beta U}$ in Eq.~(\ref{eq:G}). While
for strong e-ph coupling, the Coulomb repulsion mainly shifts the
positions of the spectrum of $G_e$ while leaving its magnitude
almost unaffected. This is because for large $\lambda$, while the resonance positions $\Delta^{(2)}_{nm}$ depend on $U$,  the
Green's functions are dominated by $D_{nm}$, which, however, is $U$ independent.

The $\lambda$ and $U$ dependence of thermal conductance $\kappa_e$
is similar to that of $G_e$, as shown in Figs.~\ref{L}(c) and~\ref{L}(d). We
note that the resonance positions of $\kappa_e$ do not coincide with
those of $G_e$ but instead coincide with the valleys of $G_e$. This
arrangement reflects the different ways in which the inelastic
scattering induced by e-ph coupling and e-e interaction degrade heat
and electrical currents. In fact, around the resonances of $G_e$, from Eq.~(\ref{Ln}) it is clear that $\mathcal L_{1,2}\simeq0$, so we have small values of $\kappa_e$ from the definition Eq.~(\ref{ke}). As a consequence, we obtain
the strong violation of WF law $L/L_0\ll1$ around the resonance
points of $G_e$ [see Figs.~\ref{L}(e) and~\ref{L}(f)].

%%%%%%%%%%%%%%%%%%%%%%%%%%%%%%%%%%%%%%%%%%%%%%%%%%%%%%%%%%%%%%%%%%%%%%%%%%%%%%%%%%%%%%%%%%%
\begin{figure}%[!htb]
\scalebox{0.38}[0.36]{\includegraphics{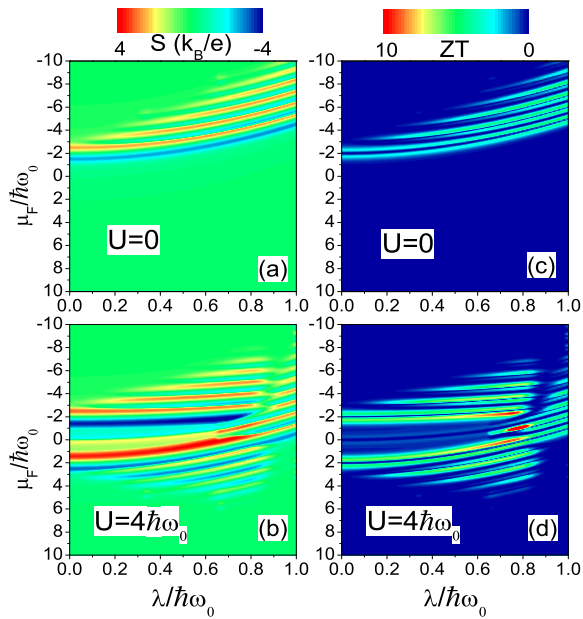}}\vspace{-3mm}
%\centering\includegraphics[width=1.0\linewidth]{ZT.pdf}
\caption{(Color online) $S$ and $ZT$ as a function of $\mu_F$ for
different $\lambda$ and $U$. The Hamiltonian parameters are the same
as those in Fig.~\ref{L}.}\label{ZT}
\end{figure}
%%%%%%%%%%%%%%%%%%%%%%%%%%%%%%%%%%%%%%%%%%%%%%%%%%%%%%%%%%%%%%%%%%%%%%%%%%%%%%%%%%%%%%%%%%%

Because $ZT=S^2/L$ becomes large as the Lorenz ratio goes to zero
while the thermopower remains finite, we expect large $ZT$ peaks
will emerge around the resonance points of $G_e$. However, as shown
in Fig.~\ref{ZT}, large $ZT$ occurs only at the sides of the
resonances and drops back to zero dramatically at exact resonance positions. This
occurs because the particle-hole symmetry at
the resonances [$\mathcal L_{1}\simeq0$ from the definition
Eq.~(\ref{Ln})] zeros the thermopower [$S\simeq0$ from the definition Eq.~(\ref{S})]. Moreover, increasing $\lambda$
increases $ZT$ as well as the number of $ZT$ peaks, which shows that
large $ZT$ is favored by multi-phonon-assisted tunneling. In
addition, the Coulomb repulsion increases $ZT$ such that optimal
$ZT$ is obtained at the merging regime of two resonant branches,
which is a consequence of the competition of e-ph coupling and
Coulomb repulsion. In other words, the optimal $ZT$ is located at $\tilde\varepsilon_{\sigma\bar\sigma}+U=\tilde\varepsilon_{\sigma}$ as we discussed above in Figs.~{\ref{L}(a) and~\ref{L}(b)}. Taking $\varepsilon_{\sigma}=\varepsilon_{\bar\sigma}=-2\hbar\omega_0, U=4\hbar\omega_0$ as in Fig.~{\ref{L}(b)}, we then can predict that the optimal $ZT$ can occur when we choose $\lambda_{\sigma}=\sqrt{2/3}\hbar\omega_0\approx 0.81\hbar\omega_0$, which is indeed the case shown in Fig.~{\ref{ZT}(d)}. In turn, given the molecular level energy and the e-ph coupling strength, we also can choose a proper e-e repulsion strength $U$ to optimize the efficiency of thermoelectricity.

Finally, the temperature dependence of $ZT$ is illustrated in Fig.~\ref{T}.
Interestingly, we see that increasing $T$ first decreases and then
increases $ZT$. Coulomb repulsion enhances $ZT$ at low temperature
but suppresses it at room temperature. Nevertheless, $ZT$ remains
large. This result indicates a great potential for
single molecular junctions as good thermoelectric devices over a
wide range of temperatures.

%%%%%%%%%%%%%%%%%%%%%%%%%%%%%%%%%%%%%%%%%%%%%%%%%%%%%%%%%%%%%%%%%%%%%%%%%%%%%%%%%%%%%%%%%%%
%%%%%%%%%%%%%%%%%%%%%%%%%%%%%%%%%%%%%%%%%%%%%%%%%%%%%%%%%%%%%%%%%%%%%%%%%%%%%%%%%%%%%%%%%%%

In view of the substantial progress that has been made in the area of molecular
devices,~\cite{NJTao:2006}
the findings in the present work will open a new avenue for using molecular junctions for the optimal
design of efficient energy conversion devices. To experimentally observe the predicted $ZT$ enhancement,
the molecular junction should be gated to tune the conduction electron density. Additionally, there is also the flexibility of
adjusting the coupling geometry between molecule and leads.

\section{Conclusions}

In conclusion, by a transformation of the phonon basis we were able
to nonperturbatively deal with the molecular quantum dot for arbitrary e-ph
coupling and e-e interaction strengths. After analytically calculating its Green's functions, we  coupled the molecular quantum dot to the electrode leads in the weak tunneling limit, and then computed the thermoelectric transport properties numerically.
We studied the synergistic effect of e-ph and e-e interactions and showed that at low
temperatures large $ZT$ occurs at the sides of resonances in
electronic conductance but drops dramatically to zero at resonance.
We found that increasing e-ph and e-e interactions increases $ZT$, although with
$G_e$ repressed. In particular, large $ZT$ is favored by multi-phonon-assisted tunneling.  More interestingly, we found that an optimal $ZT$ emerges when these two interactions were competing. Finally, we showed that a large
$ZT$ can be obtained in a wide range of temperatures.

It would be interesting to consider the Zeeman splitting of the molecular orbital energies produced, for example, by an external magnetic field or
ferromagnetic leads. Spin-related thermoelectric effects may aid the
optimal design of novel thermal-spintronic devices and
single-molecule-magnet junctions.~\cite{Dubi09Wang10} Extending our nonperturbative approach to electronic  coupling to  multiple vibrational modes~\cite{Hartle09} will be an interesting topic. It would also be desirable  to combine the present method with some {\em ab initio} electron structure theory, like the density functional theory within the local density approximations, for  more realistic calculations.~\cite{Haugbook} Finally, we would like to remark that calculating the exact self-energy $\Sigma_{\mathrm{lead}}$ remains an important open question, which deserves investigation in the future.

\begin{acknowledgments}
J.R. acknowledges the hospitality of Los Alamos National Laboratory
(LANL), where this work was carried out. J.X.Z and J.E.G.
acknowledge the support of U.S. DOE under Contract No.
DE-AC52-06NA25396. The work of J.R., C.W., and B.L. was supported in
part by NUS Grant No. R-144-000-285-646.
\end{acknowledgments}

\appendix
\section{Calculation details of nonequilibrium Green's functions}
\label{appendixA}

Before calculating the retarded (advanced) Green's function, we first detail the calculation of the lesser Green's function
$G_{\sigma}^<(t)=i\langle d^{\dag}_{\sigma}(0)d_{\sigma}(t)\rangle$
in the frequency domain:
%\begin{widetext}
\begin{align}
G_{\sigma}^<(\omega)=& i\int^{+\infty}_{-\infty}dt e^{i\omega
t}\langle
d_{\sigma}^{\dag}(0)d_{\sigma}(t)\rangle   \nonumber\\
=&i\int^{+\infty}_{-\infty}dt
e^{i\omega t}\sum_{\varphi}\sum_{\psi}\langle\varphi|\rho
d_{\sigma}^{\dag}(0)|\psi\rangle            \nonumber\\ &\times\langle\psi|e^{iH_{\mathrm{mol}}t}d_{\sigma}(0)e^{-iH_{\mathrm{mol}}t}|\varphi\rangle \nonumber\\
=& \frac{i}{Z}\int^{+\infty}_{-\infty}dt e^{i\omega
t}\sum_{\varphi}\sum_{\psi}e^{-\beta
E_{\varphi}}e^{-i(E_{\varphi}-E_{\psi})t}    \nonumber\\
&\times\langle\varphi|d_{\sigma}^{\dag}(0)|\psi\rangle\langle\psi|d_{\sigma}(0)|\varphi\rangle \nonumber\\
=& \frac{2\pi
i}{Z}\sum_{\varphi}\sum_{\psi}\delta(\omega-(E_{\varphi}-E_{\psi}))e^{-\beta
E_{\varphi}}     \nonumber\\
&\times\langle\varphi|d_{\sigma}^{\dag}(0)|\psi\rangle\langle\psi|d_{\sigma}(0)|\varphi\rangle,
\end{align}
%\end{widetext}
where we used $\rho=e^{-\beta H_{\mathrm{mol}}}/Z$ with
$Z=\text{Tr}(e^{-\beta H_{\mathrm{mol}}})=(1+N_{\mathrm{ph}})(1+e^{-\beta
\tilde\varepsilon_{\sigma}}+e^{-\beta
\tilde\varepsilon_{\bar\sigma}}+e^{-\beta
(\tilde\varepsilon_{\sigma\bar\sigma}+U)})$ and $N_{\mathrm{ph}}$ is the Bose distribution $N_{ph}=1/(e^{\beta
\omega_0}-1)$. Here $|\varphi\rangle$ and $|\psi\rangle$ are the possible
eigenstates $|\o, n\rangle_{\o}$, $|\sigma,n\rangle_{\sigma}$,
$|\bar\sigma,n\rangle_{\bar\sigma}$,
$|\sigma\bar\sigma,n\rangle_{\sigma\bar\sigma}$, and $E_{\varphi}$ and
$E_{\psi}$ are the corresponding possible eigenvalues $n\omega_0$,
$n\omega_0+\tilde\varepsilon_{\sigma}$,
$n\omega_0+\tilde\varepsilon_{\bar\sigma}$,
$n\omega_0+\tilde\varepsilon_{\sigma\bar\sigma}+U$.

There are only two nonzero combinations of $|\varphi\rangle$ and
$|\psi\rangle$ for calculating $G_{\sigma}^<(\omega)$: (1)
$|\varphi\rangle=|\sigma,n\rangle_{\sigma}$ and
$|\psi\rangle=|\o,m\rangle_{\o}$, or (2)
$|\varphi\rangle=|\sigma\bar\sigma,n\rangle_{\sigma\bar\sigma}$ and
$|\psi\rangle=|\bar\sigma,m\rangle_{\bar\sigma}$, such that the
lesser Green's function can be reduced to
\begin{align}
%\begin{eqnarray}
G_{\sigma}^<(\omega) =& \frac{2\pi
i}{Z}\sum^{\infty}_{n=0}\sum^{\infty}_{m=0}\big[\delta(\omega-(n-m)\omega_0-\tilde\varepsilon_{\sigma}) \nonumber\\
&\times e^{-\beta
(n\omega_0+\tilde\varepsilon_{\sigma})}{_\sigma\langle}
n|m\rangle_{\o}\;{_{\o}\langle} m|n\rangle_{\sigma}
\nonumber\\
&+\delta(\omega-(n-m)\omega_0-(\tilde\varepsilon_{\sigma\bar\sigma}-\tilde\varepsilon_{\bar\sigma}+U)) \nonumber\\
&\times e^{-\beta
(n\omega_0+\tilde\varepsilon_{\sigma\bar\sigma}+U)}{_{\sigma\bar\sigma}\langle}
n|m\rangle_{\bar\sigma}\;{_{\bar\sigma}\langle}
m|n\rangle_{\sigma\bar\sigma}\big].
\end{align}
%\end{eqnarray}
The detailed expression of ${_b\langle} n|m\rangle_{c}$, denoting
the inner product of modified phonon states with effective
displacements $g_b$ and $g_c$, can be derived as follows:
%\begin{widetext}
\begin{eqnarray}
{_b\langle} n|m\rangle_{c} &=&
\langle0|\frac{(\hat{a}+g_b)^n}{\sqrt{n!}}
\exp{(-{g_b}^2/2-g_b\hat{a})} \nonumber\\
&&\times\frac{(\hat{a}^{\dag}+g_c)^m}{\sqrt{m!}}
\exp{(-{g_c}^2/2-g_c\hat{a}^{\dag})}|0\rangle \nonumber\\
&=& \frac{\exp{[-(g_b-g_c)^2/2]}}{\sqrt{n!
m!}}    \nonumber\\
&&\times\langle0|(\hat{a}+g_b)^n
e^{(-g_c\hat{a}^{\dag})}e^{(-g_b\hat{a})}(\hat{a}^{\dag}+g_c)^m
|0\rangle \nonumber\\
&=& \frac{\exp{[-(g_b-g_c)^2/2]}}{\sqrt{n!
m!}}    \nonumber\\
&&\times\langle0|(\hat{a}+g_b-g_c)^n (\hat{a}^{\dag}+g_c-g_b)^m
|0\rangle \nonumber\\
&=& \frac{\exp{[-(g_b-g_c)^2/2]}}{\sqrt{n!
m!}}    \nonumber\\
&&\times\sum^{\mathrm{min}\{n,m\}}_{k=0}k! C^{k}_{n}(g_b-g_c)^{n-k}C^{k}_{m}(g_c-g_b)^{m-k} \nonumber\\
&=& (-1)^m D_{nm}(g_b-g_c)
\end{eqnarray}
%\end{widetext}
where
$$D_{nm}(x)=e^{-x^2/2}\sum^{\mathrm{min}\{n,m\}}_{k=0}\frac{(-1)^k\sqrt{n!
m!}x^{n+m-2k}}{(n-k)!(m-k)!k!}$$ is invariant under the exchange of
indices $n, m$. Note, to get the third equivalence, we utilized the
relation $\exp{(c\hat{a})}f(\hat{a}^{\dag},\hat{a})=f(\hat{a}^{\dag}+c,\hat{a})\exp{(c\hat{a})}$.

Therefore, the lesser Green's function can be further reduced:
\begin{eqnarray}
G_{\sigma}^<(\omega)= \frac{2\pi
i}{Z}\sum^{\infty}_{n,m=0}\left[\delta(\omega-\Delta_{nm}^{(1)})e^{-\beta
(n\omega_0+\tilde\varepsilon_{\sigma})}+\right. \nonumber\\
\left.\delta(\omega-\Delta_{nm}^{(2)})e^{-\beta
(n\omega_0+\tilde\varepsilon_{\sigma\bar\sigma}+U)}\right]D^2_{nm}(g_{\sigma}),
\end{eqnarray}
where
\begin{eqnarray}
\Delta_{nm}^{(1)}&=&(n-m)\omega_0+\tilde\varepsilon_{\sigma}, \\
\Delta_{nm}^{(2)}&=&(n-m)\omega_0+(\tilde\varepsilon_{\sigma}-2\omega_0g_{\sigma}g_{\bar\sigma}+U).
\end{eqnarray}
Similarly, for the greater Green's function
$G^>_{\sigma}(t)=-i\langle
d_{\sigma}(t)d^{\dag}_{\sigma}(0)\rangle$, we can obtain
\begin{eqnarray}
G_{\sigma}^>(\omega) &=& -i\int dt e^{i\omega t}\langle
d_{\sigma}(t)d_{\sigma}^{\dag}(0)\rangle  \nonumber\\
&=& -i\int dt e^{i\omega
t}\sum_{\varphi}\sum_{\psi}\langle\varphi|\rho
e^{iH_{\mathrm{mol}}t}d_{\sigma}(0)e^{-iH_{\mathrm{mol}}t}|\psi\rangle  \nonumber\\
&&\times\langle\psi|d_{\sigma}^{\dag}(0)|\varphi\rangle
\nonumber\\
&=& -\frac{2\pi i}{Z}\sum_{\varphi}\sum_{\psi}\delta(\omega+E_{\varphi}-E_{\psi})e^{-\beta
E_{\varphi}}    \nonumber\\
&&\times\langle\varphi|d_{\sigma}|\psi\rangle\langle\psi|d_{\sigma}^{\dag}|\varphi\rangle
\nonumber\\
&=&-\frac{2\pi
i}{Z}\sum_{n,m}\left[\delta(\omega-\Delta_{nm}^{(1)})e^{-\beta
m\omega_0}+\right.     \nonumber\\
&&\left.\delta(\omega-\Delta_{nm}^{(2)})e^{-\beta
(m\omega_0+\tilde\varepsilon_{\bar\sigma})}\right]D^2_{nm}(g_{\sigma}).
\end{eqnarray}
Note here the two nonzero combinations of $|\varphi\rangle$ and
$|\psi\rangle$ for calculating $G_{\sigma}^>(\omega)$ are (1)
$|\varphi\rangle=|\o,m\rangle_{\o}$ and
$|\psi\rangle=|\sigma,n\rangle_{\sigma}$, or (2)
$|\varphi\rangle=|\bar\sigma,m\rangle_{\bar\sigma}$ and
$|\psi\rangle=|\sigma\bar\sigma,n\rangle_{\sigma\bar\sigma}$.

Then, following the relation $G^r(t)=\Theta(t)(G^>(t)-G^<(t))$,
$G^a(t)=-\Theta(-t)(G^>(t)-G^<(t))$, and utilizing $$\Theta(t)=\int
\frac{d\omega}{2\pi i}\frac{e^{i\omega t}}{\omega-i0^{+}},$$ we have
the retarded (advanced) Green's function:
\begin{eqnarray}
G_{\sigma}^{r(a)}(\omega) &=&
\int\frac{d\omega_1}{2\pi}\int\frac{d\omega_2}{2\pi i}\int dt
e^{i\omega
t}\frac{e^{-i(\omega_1-\omega_2)t}}{\omega_2\mp i0^{+}}\nonumber \\ &&\times[G_{\sigma}^>(\omega_1)-G_{\sigma}^<(\omega_1)]
\nonumber \\
&=& \int\frac{d\omega_1}{2\pi
i}\frac{G_{\sigma}^>(\omega_1)-G_{\sigma}^<(\omega_1)}{\omega_1-\omega\mp i0^{+}}.
\end{eqnarray}
Substituting the expressions of the greater and lesser Green's functions, we arrive at Eq.~(\ref{eq:G}).

\section{Derivations for the current expression}
\label{appendixB}

Here we detail the calculation of the current through the
interacting system. The electronic current from left contact to
central system is defined as
$J_L=-e\langle\sum_{k\sigma}{dN^L_{k\sigma}}/{dt}\rangle$, which is
generally reexpressed as~\cite{Haugbook}
\begin{equation}
J_L=\frac{ie}{\hbar}\int\frac{d\omega}{2\pi}\sum_{\sigma}\left[\Gamma_{\sigma}^LG^<_{\mathrm{tot},\sigma}+\Gamma_{\sigma}^Lf_L(G^r_{\mathrm{tot},\sigma}-G^a_{\mathrm{tot},\sigma})\right].
\end{equation}
Substituting the expressions of various nonequilibrium Green's
functions,
\begin{eqnarray}
G^{r(a)}_{\mathrm{tot},\sigma}&=&\frac{1}{(G^{r(a)}_{\sigma})^{-1}\pm
i\Gamma_{\sigma}}, \\
G^{<(>)}_{\mathrm{tot},\sigma}&=&G^r_{\mathrm{tot},\sigma}\Sigma^{<(>)}_{\mathrm{tot},\sigma}G^a_{\mathrm{tot},\sigma},
\end{eqnarray}
we have
\begin{widetext}
\begin{eqnarray}
J_L&=&\frac{ie}{\hbar}\int\frac{d\omega}{2\pi}\sum_{\sigma}[\Gamma_{\sigma}^LG^<_{\mathrm{tot},\sigma}+\Gamma_{\sigma}^Lf_L(G^r_{\mathrm{tot},\sigma}-G^a_{\mathrm{tot},\sigma})] \\
&=&\frac{ie}{\hbar}\int\frac{d\omega}{2\pi}\sum_{\sigma}\frac{\Gamma^L_{\sigma}(\Sigma^{<}_{\mathrm{lead},\sigma}+\Sigma^{<}_{\mathrm{int},\sigma})-(2i\Gamma_{\sigma}+(G^{r}_{\sigma})^{-1}-(G^{a}_{\sigma})^{-1})\Gamma_{\sigma}^Lf_L}{[(G^{r}_{\sigma})^{-1}+i\Gamma_{\sigma}][(G^{a}_{\sigma})^{-1}-i\Gamma_{\sigma}]}
\\
&=&\frac{ie}{\hbar}\int\frac{d\omega}{2\pi}\sum_{\sigma}\frac{\Gamma^L_{\sigma}\Sigma^{<}_{\mathrm{lead},\sigma}-2i\Gamma_{\sigma}\Gamma_{\sigma}^Lf_L}{[(G^{r}_{\sigma})^{-1}+i\Gamma_{\sigma}][(G^{a}_{\sigma})^{-1}-i\Gamma_{\sigma}]}
+\frac{ie}{\hbar}\int\frac{d\omega}{2\pi}\sum_{\sigma}\frac{\Gamma^L_{\sigma}G^{<}_{\sigma}+\Gamma_{\sigma}^Lf_L(G^{r}_{\sigma}-G^{a}_{\sigma})}{[1+i\Gamma_{\sigma}G^{r}_{\sigma}][1-i\Gamma_{\sigma}G^{a}_{\sigma}]}
\\&=&\frac{e}{\hbar}\int\frac{d\omega}{2\pi}\sum_{\sigma}\frac{\Gamma^L_{\sigma}\Gamma^R_{\sigma}}{[(G^{r}_{\sigma})^{-1}+i\Gamma_{\sigma}][(G^{a}_{\sigma})^{-1}-i\Gamma_{\sigma}]}\big[f_L(\omega)-f_R(\omega)\big] \nonumber\\
&&+\frac{ie}{\hbar}\int\frac{d\omega}{2\pi}\sum_{\sigma}\frac{\Gamma^L_{\sigma}G^{<}_{\sigma}+\Gamma_{\sigma}^Lf_L(G^{>}_{\sigma}-G^{<}_{\sigma})}{[1+i\Gamma_{\sigma}G^{r}_{\sigma}][1-i\Gamma_{\sigma}G^{a}_{\sigma}]}
\end{eqnarray}
\end{widetext}
In the integration of the last line, the lesser and greater Green's
functions contain the Dirac $\delta$ functions, which only have finite
nonzero values at the resonant points. However, at those resonant
points, the retarded and advanced Green's functions have divergent
values, which finally lead to zero integration values. Therefore,
the contribution of the last integration is zero, and we finally
arrive at the electron current expression, Eq.~(\ref{eq:current}).
It is easy to get the same expression for $J_R$ with $R\leftrightarrow L$, such that current
conservation is explicitly preserved. %The present approach solves
%the current non-conserving issue $J_L\neq -J_R$ in the perturbation
%approximation of weak e-ph coupling and the canonical transformation
%for strong e-ph coupling, while in the past, people usually choose
%to calculate the symmetrized current $J\doteq(J_L+J_R)/2$ to cover
%the problem.

Following the first law of thermodynamics $dQ=dE-\mu dN$, we
have the current relation: $I_Q\equiv\dot Q=\dot E-\mu J$. Therefore,
following the similar calculation, the heat current is
straightforwardly obtained as Eq.~(\ref{eq:heatcurrent}).
We can obtain the similar expression for the heat current through
the right reservoir $I^R_Q$. %Energy conservation gives the relation $-\omega_0\dot N_{\mathrm{ph}}=I^L_Q+I^R_Q$. For linear response, the system is in equilibrium such that $\dot N_{\mathrm{ph}}=0$ and $I^L_Q=-I^R_Q$, since $N_{\mathrm{ph}}$ follows the Bose
%distribution of fixed value. Therefore, the equilibrium temperature is directly used for calculation. For the nonlinear response case, the system is in non-equilibrium and $I^{L(R)}_Q$ contains the effective inverse
%temperature $\beta_{\mathrm{eff}}$ which is a function of $N_{\mathrm{ph}}$:
%$\beta_{\mathrm{eff}}=\frac{1}{\omega_0}\ln\frac{1+N_{\mathrm{ph}}}{N_{\mathrm{ph}}}$.
%Thus, to obtain the phonon population and effective temperature, we can iteratively obtain the
%steady state solution of $0=-\omega_0\dot N_{\mathrm{ph}}=I^L_Q+I^R_Q$ which
%gives the non-equilbirium $N_{\mathrm{ph}}$ and $\beta_{\mathrm{eff}}$, and in
%turn gives non-equilibrium currents and other transport properties.
Based on the electron current expression and the expression of heat
current carried by the electron, we are capable of investigating the
thermoelectric transport properties.

\end{document}